\documentclass[vecphys]{svmult}
\usepackage[margin=1.2in]{geometry}

\usepackage{amsmath}

\usepackage{makeidx}         
\usepackage{graphicx}        
\usepackage{multicol}        
\usepackage{cite}            
\usepackage[bottom]{footmisc}

\makeindex             

\allowdisplaybreaks[2]
\begin{document}

\title{Energy current and energy fluctuations in driven quantum wires}
\author{D. Crivelli$^{1}$ \and M. Mierzejewski$^{1}$ 
\and  P. Prelov\v{s}ek$^{2}$
}
\institute{Institute of Physics, University of Silesia, 40-007 Katowice, Poland
\and
Faculty of Mathematics and Physics, University of Ljubljana;
J. Stefan Institute, SI-1000 Ljubljana, Slovenia
}

\maketitle

\begin{abstract}
We discuss the energy current and the energy fluctuations in an isolated quantum wire driven far from equilibrium. The system consists of interacting spinless fermions and is driven by a time--dependent magnetic flux. The energy current is defined by the continuity equation for the energy density which is derived both for homogeneous as well as for inhomogeneous systems. Since the total energy is not conserved in the driven system, the continuity equation includes the source terms which are shown to represent the Joule heating effects. For short times and weak drivings the energy current agrees with the linear response theory. For stronger fields or longer times of driving the system enters the quasiequilibrium regime when the energy current gradually diminishes due to the heating effects. Finally, for even stronger driving the energy current is shown to undergo a damped Bloch oscillations.       
The energy spread  also increases upon driving. However, the time--dependence of this quantity in the low field regime is quite
unexpected since it is determined mostly by the time of driving being quite independent of the instantaneous energy of the system.
\end{abstract}

\noindent

\section{Introduction}
\label{sec:1}
Understanding the nonequilibrium physics of particle and energy currents  
in correlated systems 
is important  for  various applications of novel materials. 
The present electronic and photovoltaic technologies are based 
on semiconductors, where the electron-electron interactions do not play any essential role. 
From  this point of view, the recent intensive studies on driven strongly correlated
systems\cite{matsuda1994,okamoto2010,okamoto2011,taguchi2000,oka2003,oka2005,eckstein2010,zala2012} are promising.
However, the nonequilibrium dynamics of various excitations in solids or nanosystems is usually too complex to be grasped in terms of a simple physical picture. 
Hence these studies pose a real challenge both for the experiment as well as for the theory.  

The  evolution of even the simplest quantum system is already a complicated problem 
with only few exactly solvable examples. In most cases the presence of many--body interactions 
makes this problem intractable for purely analytical approaches, hence
majority of theoretical results have  been obtained from recently developed numerical methods \cite{white2004,jim2006,my1,mierzejewski2011,prosen2010,aron2012,bukov2012,bonca2012}.  
Many studies focus on  charge dynamics in systems driven by strong electromagnetic fields \cite{jim2006,hasegawa2007,sugimoto2008,takahashi2008,
my1,my3,lev2011,lev2011_1,eckstein2011,aron2012,amaricci11,einhellinger12}.
The main motivation for this research is the
the ultrafast relaxation of photoinduced carriers revealed by the femtosecond  pump--probe spectroscopy  in various strongly correlated materials \cite{matsuda1994,dalconte12,rettig12,novelli12,okamoto2010,cortes2011,kim12}.

Interaction of strong electromagnetic fields with solids is a very complex issue 
which may involve high--energy states \cite{al-Hassanieh2008} . Hence it is mostly impossible to work with microscopic Hamiltonians which include all the relevant degrees of freedom.
Fortunately,  the highly excited carriers quickly dissipate their energy due to multiple scatterings and enter the regime which is within the reach of the standard tight--binding models. 
The dynamics of photocarriers has been intensively studied in the context of their nonradiative recombination \cite{strohmaier2010,sensarma2010,al-Hassanieh2008,dias2012,zala2012a}. Various  numerical approaches have been applied, e.g.  exact diagonalization methods \cite{takahashi2002}, time--dependent density matrix renormalization group \cite{al-Hassanieh2008,dias2012}. 
or nonequilibrium dynamical mean--field approach \cite{eckstein2013}. 

In contradistinction to the quickly developing research on the charge dynamics under far--from--equilibrium conditions, the equally important problem of the energy transport remains almost unexplored. 
In particular,  understanding of the thermoelectric phenomena in quantum systems is important for heat--to--current conversion or heat pumping in the future nanoscale devices. However,  thermoelectric properties of generic low-dimensional systems have been studied mostly within equilibrium approaches, while the nonequilibrium regime has been investigated within models of essentially noninteracting particles. First results have recently been reported in \cite{leijnse,kirchner} and \cite{sanchez2013}  for quantum dots and mesoscopic systems, respectively. In particular, the lowest order corrections to LR have been studied within scattering theory in the latter paper. 

In this work we do not address directly the problem of thermoelectricity in driven nanosystems. The aim is more modest still being directly related with the thermoelectric phenomena.  In the first part we consider a microscopic Hamiltonian of correlated spinless fermions driven by external electric field and derive the continuity equation for the energy density. This allows us to derive the energy current in the presence of external driving for either homogeneous  or inhomogeneous systems.
 In subsequent part we study the energy fluctuation in driven quantum system. While irrelevant in  solids these fluctuations may be very important in the nanoscale devices. Universality of these fluctuations has recently been shown for periodically driven systems \cite{naturepol}. Here, we demonstrate that analogous (but different in details) universality  holds true also for the case  of a steady driving.

\section{Energy current in a driven wire}
\label{sec:2}
We study a one--dimensional (1D) isolated system of interacting spinless fermions  with periodic boundary conditions. The system is driven by a time--dependent  magnetic flux  $\phi(t)$  which enters only the  
kinetic energy term of the following Hamiltonian  
\begin{align}
H&= \sum_l h_l, \label{eq3} \\
h_l&= \left( -t_h e^{i\phi} c^{\dagger}_{l+1}c_l +\mathrm{H.c.}   \right) +V \tilde{n}_{l+1}\tilde{n}_l  
+\frac{1}{2} W \left( \tilde{n}_{l-1} \tilde{n}_{l+1}
+ \tilde{n}_{l} \tilde{n}_{l+2} \right),
\label{eq4}
\end{align}
where $\tilde{n}_l=n_l -\frac{1}{2}$, $n_l= c^{\dagger}_{l}c_l$, $t_h$ is the hopping integral, 
whereas $V$ and $W$ are the repulsive interaction 
strengths for particles on the nearest and the next nearest sites, respectively. The reason behind introducing $W$ is to stay away from the integrable case  ($W=0$), which shows anomalous transport characteristics \cite{my1,tomaz2011,Marko2011,Robin2011}.

The aim of studies discussed in the present section is to derive the continuity equations for particle and energy densities in the presence of external driving.  However, for the sake of completeness we start with a rather straightforward derivation of particle  and energy currents ($j^N_l$ and   $j^E_l$,
respectively) for the time--independent Hamiltonian. In the absence of driving, i.e. for a constant magnetic flux $\phi$, the particle number and the total energy are conserved, hence one  derives the continuity equations which do not contain any source terms. In the Heisenberg picture the equation of motion for
the particle density operator $n_l$ reads
\begin{equation}
\frac{\mathrm d}{{\mathrm d} t} n_l +i[n_l,H]=0,  
\label{eq1}
\end{equation}       
and the corresponding current operator $j^N_l$ fulfills 
\begin{equation}
\nabla j^N_l \equiv j^N_{l+1}-j^N_{l}=i[n_l,H].   
\label{eq2}
\end{equation} 
The solution of Eq. \ref{eq2} for the Hamiltonian   \ref{eq4}
\begin{equation}
j^N_l= i t_h \exp(i\phi) c^{\dagger}_{l+1}c_{l}+ \mathrm{H.c.},
\label{jn}
\end{equation}
fulfills also the well known relation $j^N_l=-\frac{{\mathrm d} h_l}{{\mathrm d} \phi}$.
In order to determine the energy current we have defined  the energy density $h_l$. Since $H$ can be split into $h_l$ in many inequivalent ways, the energy current operator is not uniquely defined either. In Eq. (\ref{eq4}) we take $h_l$ which has support symmetric with respect to the bond between sites $l$ and $l+1$.  Then, similarly to Eq. (\ref{eq2}), one 
defines the energy current through the continuity equation as
\begin{equation}
\frac{\mathrm d}{{\mathrm d} t} h_l+ i[h_l,H]=\frac{\mathrm d}{{\mathrm d} t} h_l+j^E_{l+1}-j^E_{l}=0.
\label{jedef}
\end{equation} 
The calculations are straightforward but tedious. 
For a  translationally invariant system one usually
considers the current  
averaged over the whole system $j^E=\frac{1}{L} \sum_l j^E_{l}$:
\begin{align}
j^E =& \,\frac{1}{L}\sum_l (-t_h^2) [i \exp(2i\phi) c^{\dagger}_{l+1}c_{l-1}+{\mathrm H.c.}] \nonumber  \\
 & + \frac{1}{L} \sum_l j^N_l \left[\frac{3W}{2} (\tilde{n}_{l+3}+\tilde{n}_{l-2})+ \frac{2V-W}{2} (\tilde{n}_{l+2} 
 +\tilde{n}_{l-1}) \right].
 \label{je}
\end{align}
 \textit{The energy current in a driven  system}. For the time--dependent Hamiltonian the energy is not  conserved
hence the continuity equation for $h_l$ may include source terms. 
Other important difference with respect to the previous case
is that now it is easier to carry out calculations, at least initially, in the Schr\"odinger picture. For arbitrary
$| \psi_t \rangle$ and $ | \xi_t  \rangle$ one finds directly from the Schr\"odinger equation  
\begin{align}
\frac{\mathrm d}{{\mathrm d} t}  
\langle \psi_t |h_l(t) | \xi_t  \rangle +i\langle \psi_t |[h_l,H]| \xi_t  \rangle = 
\langle \psi_t |
\frac{{\mathrm d} h_l}{{\mathrm d} \phi} | \xi_t  \rangle\, \dot{\phi}  , && \\
\frac{\mathrm d}{{\mathrm d} t}  
\langle \psi_t |H(t) | \xi_t  \rangle = 
  \langle \psi_t |
\frac{{\mathrm d} H}{{\mathrm d} \phi} | \xi_t  \rangle\, \dot{\phi},
\end{align} 
Using Eqs. (\ref{eq4}-\ref{jedef}) and introducing the
time--dependent electric field $F(t)=-\dot{\phi}(t)$ one gets
\begin{align}
\frac{\mathrm d}{{\mathrm d} t}  
\langle \psi_t |h_l(t) | \xi_t  \rangle +
\langle \psi_t  | \nabla j^E_l | \xi_t  \rangle = 
F  \langle \psi_t | j^N_l | \xi_t  \rangle, & \label{cont} \\
\frac{\mathrm d}{{\mathrm d} t}  
\langle \psi_t |H(t) | \xi_t  \rangle = 
F  \langle \psi_t | {\textstyle \sum_l} j^N_l | \xi_t  \rangle, &\label{conttest} 
\end{align} 
where both current operators are defined without driving. 
The main issue is to set whether the
term at the rhs. of Eq. (\ref{cont}) represents the source
of energy or whether it should be accommodated into a new 
current operator $\nabla \bar{j}^E_l $.  In the latter scenario
one would end up with the continuity equation 
$\frac{\mathrm d}{{\mathrm d} t}  
\langle \psi_t |h_l(t) | \xi_t  \rangle +
\langle \psi_t  | \nabla \bar{j}^E_l | \xi_t  \rangle =  0 $, which
for periodic boundary conditions implies conservation of the total
energy. The latter result follows from the identity $\sum_l \nabla \bar{j}^E_l=\sum_l (\bar{j}^E_{l+1}-\bar{j}^E_{l}) =0$ which holds for
any $\bar{j}^E_{l}$.  However conservation of the total 
energy would violate Eq. (\ref{conttest}). Consequently, the nonequilibrium term $F j^N_l$  is a source of energy, while the energy current operator remains the same as for the case without driving.  Note, that this reasoning holds true independently of any particular form of $h_l$.

For $| \xi_t  \rangle = | \psi_t \rangle$,
Eq. (\ref{cont}) turns into the continuity equation for the
expectation value  $  \langle h_l(t)  \rangle  =\langle \psi_t |h_l(t) | \psi_t  \rangle$:
\begin{equation}
\frac{\mathrm d}{{\mathrm d} t}  
\langle h_l \rangle +
\nabla \langle j^E_l  \rangle = 
F  \langle j^N_l \rangle
\end{equation}
The same continuity equation may be derived for a system in a mixed state  when $ \langle h_l  \rangle  = {\mathrm Tr }[\rho(t) h_l(t) ]$ and the density matrix
$\rho(t)$ evolves according to the von Neumann equation. Still, it might seem disturbing that the continuity equation for the driven
case concerns the expectation values, while Eq. (\ref{jedef}) has been derived entirely in the operator language. The evolution of an isolated system (whether driven or not) is unitary.
Putting $| \psi_t  \rangle = U(t,t_0) | \psi_{t0} \rangle$
and $| \xi_t  \rangle = U(t,t_0) | \xi_{t0} \rangle$ into
Eq. (\ref{cont}) and recalling that this equation holds
for arbitrary $ | \psi_{t0} \rangle$ and $ | \xi_{t0} \rangle$
one finds the continuity equation
\begin{equation}
\frac{\mathrm d}{{\mathrm d} t}  
\tilde{h}_l(t) +
 \nabla \tilde{j} ^E_l = 
F  \tilde{j}^N_l,
\end{equation}
where the operators with tilde are defined through 
the unitary transformation 
$\tilde{h}_l(t) = U^{\dagger}(t,t_0)  h_l(t) U (t,t_0)$ and
$U^{\dagger}(t,t_0) U (t,t_0)=1$. While the continuity equation
can be written in the operator language also for driven systems,
in most cases this form is rather useless because of complicated form
or the evolution operator $U (t,t_0)$.  
  
Finally we turn to the most general case of an inhomogeneous wire. 
For this reason we consider site--dependent interactions $V \rightarrow V_l$  $W\rightarrow W_l$ as well as local potentials $\varepsilon_l$. 
The energy  density takes the form:
\begin{align}
h_l&=h^{tV}_{l,l+1}+\frac{1}{2}h^{W}_{l-1,l+1}+\frac{1}{2} h^{W}_{l,l+2}\\
h^{tV}_{l,l+1} &= \left( -t_h e^{i\phi} c^{\dagger}_{l+1}c_l +\mathrm{H.c.} \right) + V_l\, \tilde{n}_l \,\tilde{n}_{l+1} + \frac{1}{2} \varepsilon_{l}\, \tilde{n}_l + \frac{1}{2} \varepsilon_{l+1}\, \tilde{n}_{l+1}\\
h^{W}_{l-1,l+1} &= W_{l-1}\, \tilde{n}_{l-1} \, \tilde{n}_{l+1} 
\label{inhom_hamiltonian}
\end{align} 
This form of the local energy density has a symmetric support on sites $l-2$ through $l+2$. The partition in 3 distinct terms has been made to ease the calculation of the commutators.

From Eq. (\ref{jedef}) it is evident that we need to compute the commutator of $h_l$ with $H$ and break the term $j^E_{l+1}-j^E_{l}$ into distinct contributions to the energy current. 
\begin{align}
\frac{d}{dt} h_l = &i [H, h_l] = i\, [\,{\textstyle \sum_j h_j}, h_l]\\                    
                 = &i [h_{l-3} + h_{l-2} + h_{l-1} + h_{l} + h_{l+1} + h_{l+2}+h_{l+3}, 
						h_l]
\end{align}
since all terms with $|l-j| \geq 4$ share no common operators and commute. Writing explicitly the values for all $h_l$,
\begin{align}
\frac{d}{dt} h_l = &-i\, [ h^{tV}_{l,l+1}+\frac{1}{2} h^{W}_{l-1,l+1}+\frac{1}{2}h^{W}_{l,l+2} \;,\; \\
   &+ h^{tV}_{l-3,l-2}+h^{tV}_{l-2,l-1}+h^{tV}_{l-1,l}+h^{tV}_{l+1,l+2}\\
   &+h^{tV}_{l+2,l+3}+h^{tV}_{l+3,i+4}+\frac{1}{2}h^{W}_{l-4,l-2}+ h^{W}_{l-3,l-1}\\
   &+ \frac{1}{2} h^{W}_{l-2,l}
   + \frac{1}{2}h^{W}_{l-1,l+1}+ \frac{1}{2}h^{W}_{l,l+2}+ h^{W}_{l+1,l+3}]
\end{align}
and expanding the big commutator, one obtains two nonzero terms involving only $h^{tV}$:
\begin{align}
\left[h^{tV}_{l,l+1},h^{tV}_{l-1,l}\right]\;,\; \left[h^{tV}_{l,l+1},h^{tV}_{l+1,l+2}\right]
\label{onlyV}
\end{align}
all terms involving only $h^W$ commute, leaving 12 nonzero mixed terms:
\begin{align}
2\, &\frac{1}{2} \left[h^{tV}_{l,l+1},h^{W}_{l-2,l}\right] \;,\; 
&2\, \frac{1}{2}\left[h^{tV}_{l,l+1},h^{W}_{l+1,l+3}\right] \; ,  \; \nonumber \\
&\frac{1}{2}\left[h^{tV}_{l,l+1},h^{W}_{l-1,l+1}\right] \;,\;
&\frac{1}{2}\left[h^{tV}_{l,l+1},h^{W}_{l,l+2}\right] \; ,\; 
&\frac{1}{2}\left[h^{W}_{l-1,l+1},h^{tV}_{l-2,l-1}\right]\;,\; \nonumber \\
&\frac{1}{2}\left[h^{W}_{l-1,l+1},h^{tV}_{l-1,l}\right]\;,\; 
&\frac{1}{2}\left[h^{W}_{l-1,l+1},h^{tV}_{l+1,l+2}\right] \; , \; 
&\frac{1}{2}\left[h^{W}_{l,l+2},h^{tV}_{l-1,l}\right] \; ,\; \nonumber \\
&\frac{1}{2}\left[h^{W}_{l,l+2},h^{tV}_{l+1,l+2}\right]\;,\;
&\frac{1}{2}\left[h^{W}_{l,l+2},h^{tV}_{l+2,l+3}\right]  \;,\; \nonumber
\end{align}
where the first two terms are to be counted twice in order to pair each commutator uniquely.
Before calculating the explicit values for the above operators, it is useful to separate parts of the ansatz (\ref{jedef}) with $j^E_{l} = \sum_k j^{E_k}_{l}$, leaving the task of regrouping the commutators in order to define all $j^{E_k}_{l}$. In this case, there are $14/2=7$ such contributions to be found.

The structure of Eq. (\ref{onlyV}) allows one to immediately recognize their sum as a difference between operators defined on two contiguous sites
\begin{align}
i \left[h^{tV}_{l,l+1},h^{tV}_{l+1,l+2}\right] &+ i \left[h^{tV}_{l,l+1},h^{tV}_{l-1,l}\right]   = \nonumber \\
i \left[h^{tV}_{l,l+1},h^{tV}_{l+1,l+2}\right] &- i\left[h^{tV}_{l-1,l},h^{tV}_{l,l+1}\right]  = \nonumber \\
j^{E_1}_{l+1} &- j^{E_1}_{l}
\label{je1}
\end{align}
we thus define the first current $j^{E_1}_{l}$ and look for a similar pattern, which holds for 5 of the 7 pairs. The remaining ones encode a difference between second neighbors. 
\begin{align*}
i\frac{1}{2}\left[h^{W}_{l,l+2},h^{tV}_{l+2,l+3}\right] &+ 
i\frac{1}{2}\left[h^{tV}_{l,l+1},h^{W}_{l-2,l}\right] =\\ \nonumber
i\frac{1}{2}\left[h^{W}_{l,l+2},h^{tV}_{l+2,l+3}\right] &- 
i\frac{1}{2}\left[h^{W}_{l-2,l},h^{tV}_{l,l+1}\right] =\\ \nonumber
\bar{j}^{E}_{l+2} &- \bar{j}^{E}_{l} 
\end{align*}
The double difference needs to be interpreted as arising from a partial cancellation:
\begin{align*}
\bar{j}^{E}_{l+2} - \bar{j}^{E}_{l} = (\bar{j}^{E}_{l+2} + \bar{j}^{E}_{l+1}) - (\bar{j}^{E}_{l+1} + \bar{j}^{E}_{l}),
\end{align*}
and the contribution to the current for site $l$ is taken as
\begin{align*}
j^{E_6}_{l} = \bar{j}^{E}_{l+1} + \bar{j}^{E}_{l} &= i\frac{1}{2}\left[h^{W}_{l-1,l+1},h^{tV}_{l+1,l+2}\right] + i\frac{1}{2}\left[h^{W}_{l-2,l},h^{tV}_{l,l+1}\right]
\end{align*}
The full list of currents contributing to $j^E_l$ is:
\begin{align*}
 j^{E_{1}}_{l} & =  \, i \,\left[h^{tV}_{l-1,l},h^{tV}_{l,l+1}\right] \label{jelist} \\
 j^{E_{2}}_{l} & =  \frac{1}{2} \, i \,\left[h^{tV}_{l-1,l},h^{W}_{l-1,l+1}\right] \\
 j^{E_{3}}_{l} & =  \frac{1}{2} \, i \,\left[h^{tV}_{l-1,l},h^{W}_{l,l+2}\right] \\
 j^{E_{4}}_{l} & =  \frac{1}{2} \, i \,\left[h^{W}_{l-2,l},h^{tV}_{l,l+1}\right] \\
 j^{E_{5}}_{l} & =  \frac{1}{2} \, i \,\left[h^{W}_{l-1,l+1},h^{tV}_{l,l+1}\right] \\
 j^{E_{6}}_{l} & =  \frac{1}{2} \, i \,\left[h^{tV}_{l-2,l-1},h^{W}_{l-1,l+1}\right]+\frac{1}{2} \, i \,\left[h^{tV}_{l-1,l},h^{W}_{l,l+2}\right] \\
 j^{E_{7}}_{l} & =  \frac{1}{2} \, i \,\left[h^{W}_{l-2,l},h^{tV}_{l,l+1}\right]+\frac{1}{2} \, i \,\left[h^{W}_{l-1,l+1},h^{tV}_{l+1,l+2}\right]
\end{align*}
The operators thus defined are automatically Hermitian, since they are the commutator of two Hermitian operators multiplied by $i$. The commutators are straightforward to calculate, and follow the pattern of an expression involving the number operators $\tilde{n}_{l}$ and the particle current defined in Eq. (\ref{jn}). The current term from Eq. (\ref{je1}) deviates from the rule and includes a hopping term between second neighbors.
%
We summarize all the contributions in their full functional form:
\begin{align*}
 j^{E_{1}}_{l} & =  -t_h^2 \left( i e^{2 i \phi } c^{\dagger }{}_{l+1} c_{l-1} + H.c. \right) \\
   &+ \left(\tilde{n}_{l+1} V_l + \frac{\varepsilon_{l}}{2}\right) \,j^N_{l,l-1} + \left(\tilde{n}_{l-1} V_{l-1}+\frac{\varepsilon_{l}}{2}\right) \,j^N_{l+1,l}\\
 j^{E_{2}}_{l} & =  -\frac{1}{2} \tilde{n}_{l+1} W_{l-1} \,j^N_{l,l-1} \\
 j^{E_{3}}_{l} & =  \frac{1}{2} \tilde{n}_{l+2} W_l \,j^N_{l,l-1} \\
 j^{E_{4}}_{l} & =  \frac{1}{2} \tilde{n}_{l-2} W_{l-2} \,j^N_{l+1,l} \\
 j^{E_{5}}_{l} & =  -\frac{1}{2} \tilde{n}_{l-1} W_{l-1} \,j^N_{l+1,l} \\
 j^{E_{6}}_{l} & =  \frac{1}{2} \left(\tilde{n}_{l+1} W_{l-1} \,j^N_{l-1,l-2}+\tilde{n}_{l+2} W_l \,j^N_{l,l-1}\right) \\
 j^{E_{7}}_{l} & =  \frac{1}{2} \left(\tilde{n}_{l-2} W_{l-2} \,j^N_{l+1,l}+\tilde{n}_{l-1} W_{l-1}\, j^N_{l+2,l+1}\right)
 \end{align*}
For a homogeneous, translationally invariant system, 
the average current $j^{E} = \frac{1}{L} \sum_{l,k} j^{E_k}_{l}$
reduces to Eq. (\ref{je}), with an additional $(-\varepsilon\, j^N_{l+1,l})$ contribution due to a shift of the energy by $\varepsilon$.
%

\section{Results and discussion}

Using the microcanonical (MC) Lanczos method \cite{mclm}
we generate an approximate initial state $|\Psi(0) \rangle$
with imposed energy $E_0=\langle \Psi(0)| [H(0)|\Psi(0) \rangle$ but also with a small energy uncertainty 
$\delta^2 E_0=\langle \Psi(0)| [H(0)-E_0]^2|\Psi(0) \rangle $.  
Typically, we take $L=24$ or $26$  sites and $\delta \simeq 0.01$. As required for the MC ensemble, the energy window 
is small on macroscopic scale ($\delta E_0/E_0 \ll 1$) but still contains a large number of levels. 
The initial inverse temperature 
$\beta$ can be estimated from the initial value of the kinetic energy $E^k$ from the high--temperature expansion (HTE) for the canonical ensemble. In particular, for a system
of $N$ fermions on $L$ sites the HTE gives the kinetic energy
\begin{equation}
E^k= -\frac{2\beta N (L-N)}{L-1}.
\label{beff}
\end{equation}

Then, at time $t=0$ the electric field is switched on and  
the time evolution $ |\Psi(0) \rangle \rightarrow |\Psi(t) \rangle$ 
is calculated in small time increments $\delta t < 1$  by step--vise change of $\phi(t)$.
Lanczos propagations method \cite{lantime} is applied to each time interval $(t,t+\delta t)$. An obvious restriction imposed on the time of evolution 
is to stay  within the time--window $t< \hbar / \delta E_0$, while the time--resolution $\delta t \ll \frac{1}{F}$ is dictated by the need to approximate the Hamiltonian as constant through any $\delta t$. 

We start with the energy current driven by a constant electric field in a homogeneous wire. Due the particle--hole symmetry $\langle j^E \rangle$ vanishes for the half--filled case, i.e., for $\langle \tilde{n}_l \rangle=0$. Therefore, when studying the energy current we consider a system consisting of $L=24$ sites with $N=10$ fermions. Further on, time will be expressed in units of $\hbar/t_h$ while $t_h$ will be set as the unit of energy.

\begin{figure}[t]
\centering
\includegraphics*[width=.7\textwidth]{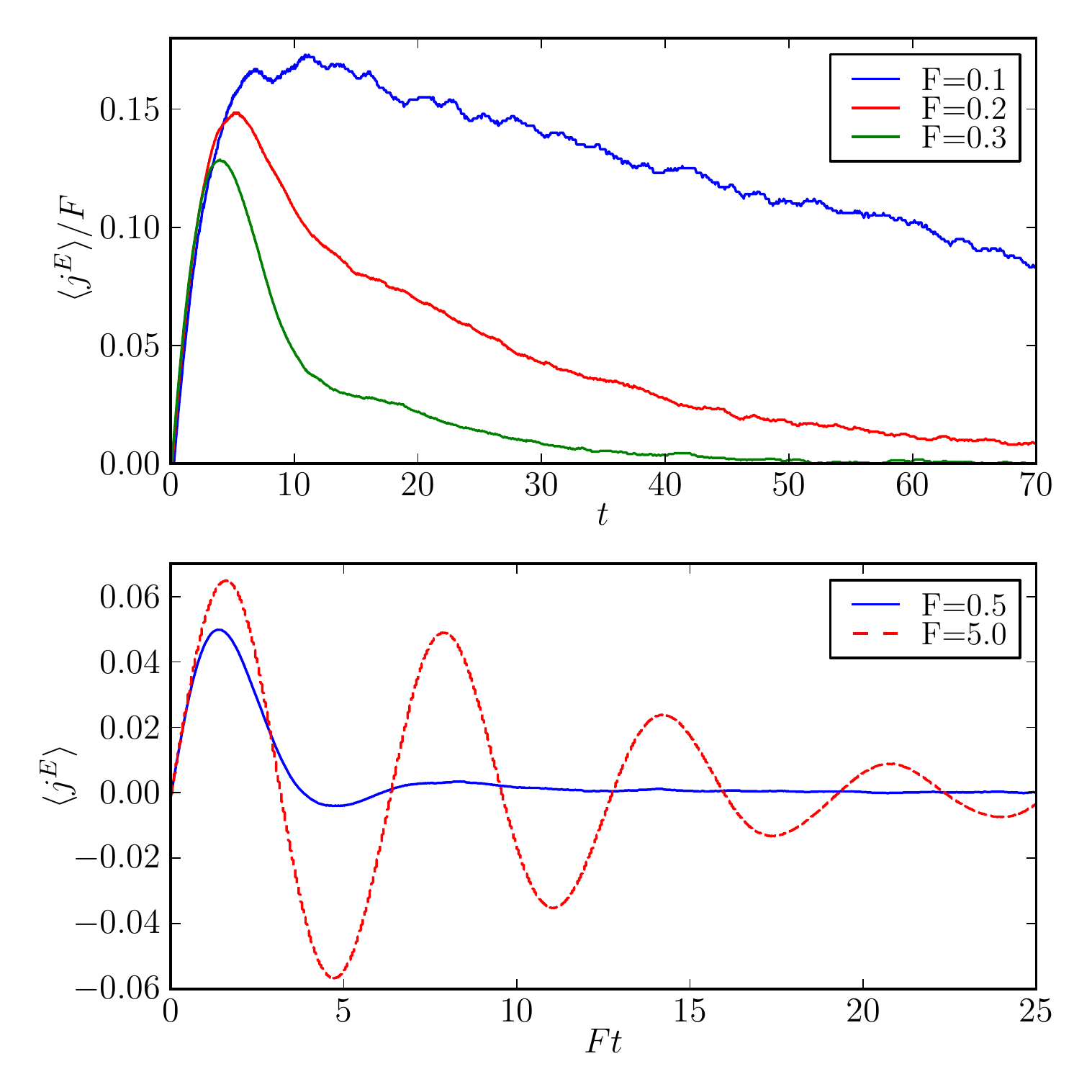}
\caption[]{Time--dependence of the energy current for various 
electric field $F$ displayed in the legend.The system consists of $N=10$ fermions on $L=24$ sites. Under equilibrium conditions, the interactions $V=1.5$, $W=1$ correspond to a metallic state at any temperature.
The results have been obtained for initial inverse temperature $\beta=0.3$. }
\label{fig:1}       
\end{figure}  

Fig. \ref{fig:1} shows the time--dependence of the energy current
$\langle j^E \rangle $ in a wire driven by low--to--moderate fields
(upper panel) as well as in the strong field regime (lower panel).
One can see that the ratio $\langle j^E \rangle/F $ is independent of the driving field at the initial stage of the evolution. It is a clear hallmark of the linear--response  (LR) regime   which always occurs for a sufficiently short time of driving. The stronger the field the sooner  $\langle j^E \rangle$ departs from the predictions of the LR theory. It is not easy to obtain the $dc$ LR directly from the real--time calculations, since for any finite $F$ the long--time regime is always beyond the LR theory.  The initial slope of the energy current is $\frac{\mathrm d}{\mathrm d t} \langle j^E \rangle_{t=0}= 
 -F \langle \partial_{\phi} j^E \rangle_{t=0} $. It is interesting to
note that $\langle \partial_{\phi} j^E \rangle_{t=0}$ is a kind of correlated hopping and represents the sum rule for LR in the initial
equilibrium state.

For longer times and/or stronger $F$, the energy current diminishes  
and eventually vanishes. In this regime one finds a counterintuitive   
result when the energy current is larger when $F$ is weaker. Similar observation has previously been found for the particle current\cite{mierzejewski2011a} and explained as the result of the Joule--heating.  As follows from Eq. (\ref{conttest}) driving with electric field increases the energy of the system. This effect 
is beyond the LR theory hence it must be at least of the order of $F^2$.
As soon this heating effect becomes visible it strongly depends on the magnitude of driving. Consequently two systems driven within the same time--window by different $F$ have exceedingly different energies. The system driven by weaker $F$ may be much {\em colder} hence it responds much stronger to the external driving.

\begin{figure}[t]
\centering
\includegraphics*[width=.7\textwidth]{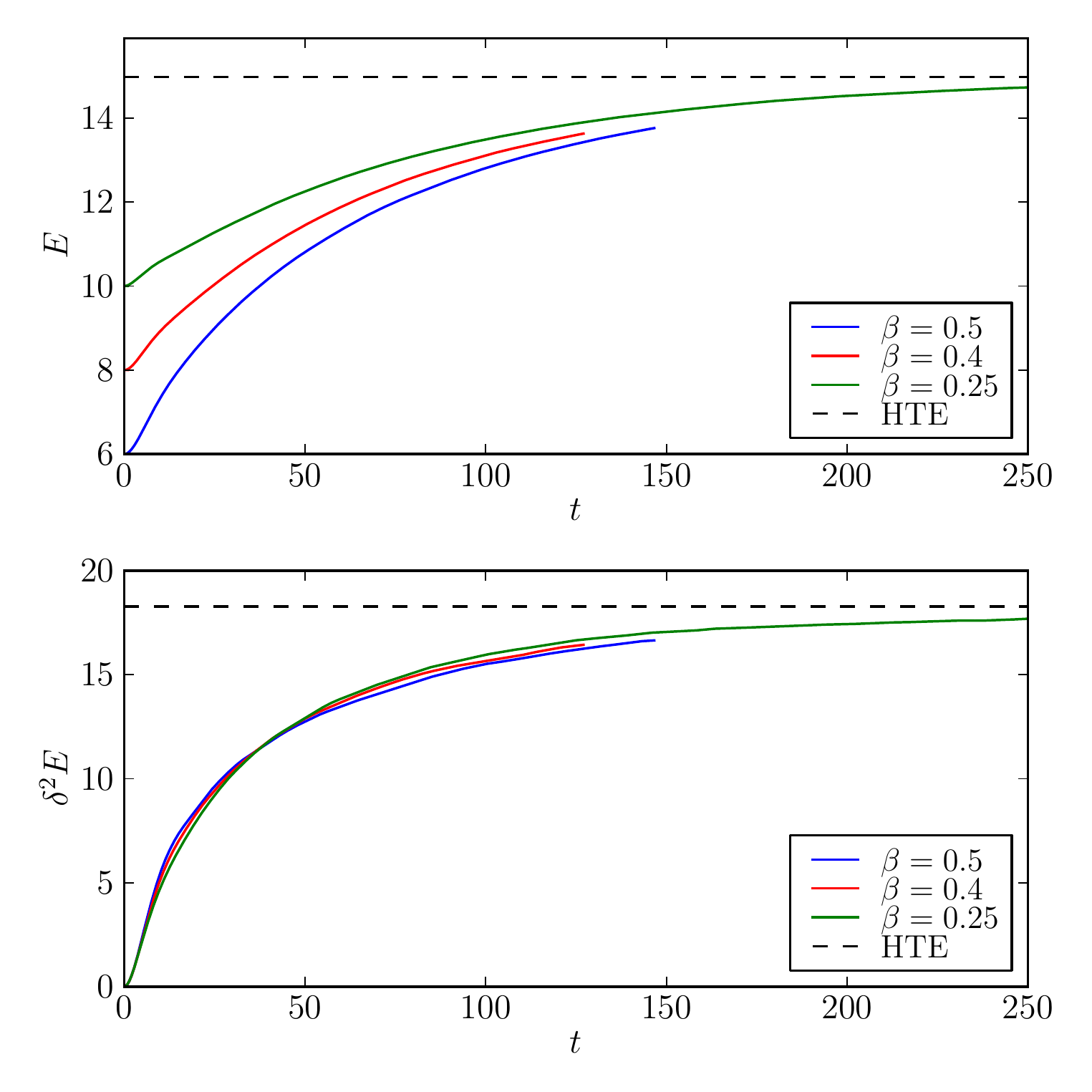}
\caption[]{Time--dependence of the average energy $E(t)$ (upper panel) as well as the
energy spread  $\delta^2 E$ 
(lower panel) for a system of $N=13$ fermions on $L=26$ sites.
Here, $V=1.4$, $W=1$ and $F=0.1$ while the initial inverse temperature
is indicated in the legend. The horizontal lines show the equilibrium 
HTE results  for infinite temperature.  
 }
\label{fig:2}       
\end{figure}

 The time--dependence of the energy current
becomes  very different for even stronger fields  as it is shown in the lower panel of Fig. \ref{fig:1}.
Namely, $\langle j^E \rangle$ starts to oscillate and this oscillations share several common features with the well known Bloch oscillations of the particle current \cite{my1}. Namely, the frequency of these oscillations is determined by the electric field while the initial amplitude of the oscillations is $F$--independent.   
  
Finally, we discuss the time--dependence of the average energy $E(t)=\langle H(t)  \rangle$ and the energy spread  $\delta^2 E =\langle  [H(t)-E(t)]^2 \rangle $. Results shown in  Fig. \ref{fig:2} have been obtained for a system driven by a weak field. Initially, the 
system is in MC state, hence $\delta E \rightarrow 0$. Upon driving,  $E(t)$ asymptotically approaches the energy of the system described by a the
canonical ensemble with $\beta=0$. Similarly to this, also the energy spread asymptotically approaches its canonical value.
However, the evolution of $\delta E$ in the low field regime is rather
unexpected since it is determined mostly by the time of driving being quite independent of the instantaneous energy of the system. Such behavior contrasts with the quasiequilibrium evolution of many local observables \cite{mierzejewski2011a} in the regime of low electric field.
Although, the instantaneous values of the latter quantities 
change in time they are determined mostly by the instantaneous energy.
Moreover, their expectation values  are close to the equilibrium result for such ensemble that $\langle H \rangle_{equilibrium}=E(t)$. 
Except from the initial MC state and the asymptotic canonical one,
 $E(t)$ and $\delta E(t)$ are independent of each other excluding the quasiequilibrium evolution of the latter quantity.
Although  the behavior of the energy spread is irrelevant 
for macroscopic systems, it might be quite important for driven nanosystems, where the ratio $\delta E/ E$ is non--negligible.

\acknowledgement  
This work has been carried out within 
the NCN project "Nonequilibrium dynamics of correlated
quantum systems". D.C. acknowledges a scholarship from the FORSZT project, co-funded by the European Social Fund.

\bibliography{mierzejewski.bib}



\printindex
\end{document}